# Defense against Lion Attack in Cognitive Radio Systems using the Markov Decision Process Approach

Khadijeh Afhamisisi

afhami@iust.ac.ir

Hadi Shahriar Shahhoseini

hshsh@iust.ac.ir

Ehsan Meamari

ehsanmeamari@gmail.com

Electrical Engineering Department, Iran University of Science and Technology

Tehran, Iran, 1684613114, Tel: +98 21 77240540

*Abstract*—Cognitive Radio (CR) technology is a solution to solve the lack of spectrum by allowing the secondary user to use licensed bands. There are several potential security challenges for cognitive radio like Jamming, PUE and Lion attack. Lion attack is multi layer attacks that has effected on two layers. The Lion attack uses PUE or jamming attack in physical layer to disrupt TCP protocol in transport layer. Since transport layer is unaware of physical layer situation, when it occurs to an unacknowledged packet, there is no way to distinguish between congestion and disconnection in the physical layer. So the windows size of TCP would be decrease because of a wrong decision caused by unawareness. To Mitigate the Lion attack the cross layer design is usually used to freeze its windows size during frequency handoff. The main issue in this solution is finding the best strategy for freezing. In this paper, we propose a dynamic method for freezing and performing frequency handoff to counter Lion attack. Also a learning model based on Markov decision process (MDP) is applied. By using the proposed method, the secondary user can choose an optimal strategy in both physical and transport layers to reduce the effect of Lion attack.

## I. INTRODUCTION

Cognitive radio network (CRN) is a new technology that enables the usage of the unused licensed spectrum bands [1]. In a cognitive radio network, the unlicensed users (secondary users) are allowed to access to the licensed bands on a non-interference basis to legacy spectrum holders (primary users).

There are several potential attacks on cognitive radio systems. Lion attack, which one of them, can be implemented by both jamming and primary user emulation (PUE) attacks to reduce the throughput of Transmission Control Protocol (TCP) by forcing frequency handoffs [2]. In Jamming attack, the malicious user intentionally transmits signal in a licensed band and jams the secondary user to reduce the signal to noise ratio. The malicious users can run a PUE attack and transmit spurious signals which emulate signals of the primary users to prevent the secondary users from accessing the network.

Lion attack is multi layer attack that uses vulnerabilities in the physical and transport layers to reduce throughput of TCP protocol. TCP protocol is used as a fixed technology for wireless and wired networks. In TCP protocol, any losses are assumed to be due to congestion; but this is not true in wireless networks, because the wireless links have higher bit error rates compared with wired networks and there are a lot of temporary disconnections for frequency handoff. Generally when the TCP acknowledge has not received. TCP protocol will assume that there is congestion and thus decrease the window size of transmitter to a minimum size. Afterward, the transmitter uses a slow start mechanism to increase the windows size. Thus a small number of packets will be sent into the link. In Lion attack, the attackers use this fact to reduce the throughput of TCP protocol by forcing the secondary user to do frequency handoff through Jamming or PUE attacks [1].

CRNs perform at the physical link layer and target to reduce the throughput of TCP. TCP needs more information than those prepared by ISO model to detect this attack. In this paper, a cross-layer architecture is used to confront the Lion attack. In the proposed cross layer architecture, the transport layer should be aware of physical layer status, to distinguish between congestion and misbehavior there. Meanwhile, we employ Markov Decision Process (MDP) as a mathematical framework for modeling the decision-making to find an optimal defense strategy against Lion attack.

There are some solutions to confront this attack. Almost all of them use cross-layer architecture to inform TCP layer from the physical layer, it can freeze the windows size of TCP protocol during frequency handoff. Although these solutions help to reduce the effects of Lion attack but they don't pay attention dynamic nature of the environment in coordinated attacks. Sometime, the secondary user can stay and freeze the windows size because of increasing the number of malicious users. The strategy of the secondary users for performing frequency handoff depends on environment parameters like number of the malicious users, the occupation probability of the band with primary user. The secondary user must observe the environment and choose a strategy for hopping or remaining and transport layer can use freezing or minimizing the window size. These decisions are related to status of environment and the secondary user should learn how to do it. In this paper, we proposed a learning method to help the secondary user to choose his/her strategy in different situation.

The rest of this paper is organized as follows. The next section presents related works in the literature. In section III, the proposed model is described for Lion attack. In section IV, MDP applied to the problem by defining states, action, rewards and determining transition probabilities. Analytical results are discussed in Section V. Section VI concludes the paper.

## II. RELATED WORK

In this section, first we review PUE and Jamming attack which are base of the Lion attack, and then describe the main lack in the prediction counter attacks of Lion which is missing strategy in finding the windows size in TCP.

In PUE attack, by forcing frequency handoff, the attacker has effects on TCP. In this case, if the attacker knows or can guess some of the connection parameters, it can perform a PUE attack by emulating a primary transmission at predicted time instants and force the secondary user to make handoff frequency. In PUE attack, the malicious users modify the air interface to mimic characteristics of the primary users' signal. Thus the secondary users deem it as the primary users' signal and leave the band [4]. This attack was first introduced by Chan and Park [5]. The main counter attack to PUE is checking the power of signals. In [6], the received powers from both primary and malicious users are compared according an analytical model. A Bayesian game is applied in [7] to model the received power from both malicious and secondary users. Li and Han studied a PUE attack by a Dogfight game in cognitive radio networks in [8-9]. They suggested the secondary users must switch to appropriate bands to defend against PUE attack. They derived the optimal switching probabilities for both secondary and malicious users can be calculated.

The Jamming attack is another attack in CRN in which the malicious users interrupt communication sessions of the secondary users by making interference. If the interference generated by malicious users is high enough, they can substantially reduce the communication performance or even completely block the secondary users [11]. Frequency hopping is an useful and simple defense method, proposed for defending against the Jamming attack, in multichannel CRNs [12]. A Markov Decision Process (MDP) model is also proposed as a countermeasure for Jamming attacks in [13].

The Lion attack is an attack which was introduced by Hern in [1]. In Lion attack, the attacker use PUE or Jamming to cause delay in physical layer result in decreasing windows size of TCP. There are several methods to prevent the Lion attacks. Le et.al [18] introduced a cross layer design by using a mobility detection element in the network layer that informs the transport layer to optimize TCP operations. TCP-Feedback has proposed in [19] to help the TCP sender distinguish between losses due to routes failure and network congestion. TCP-freeze is proposed in [20] to fit the windows size when there is no congestion. In this method the receiver informs the sender about situation of physical layer by sending ZWP signal. This approach is not suitable in dynamic environment, such as CRN, where their parameters have effect on the environment. In this situation, the secondary user finds and applies the best strategy against the malicious users.

### III. SYSTEM MODEL

Suppose there is a network with $M$ spectrum bands and one secondary user who tries to access the network. Based on IEEE 802.22, the secondary user can send information when the primary users are absent [10].

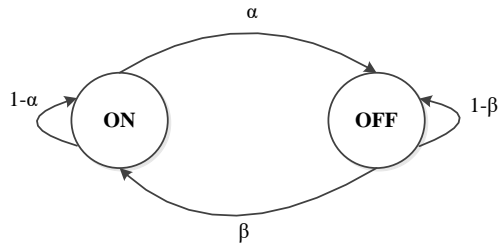

**Figure (1).the access pattern of the primary users**

The access pattern of primary users is presented by ON-OFF model as shown in Figure (1), in which switching from ON and OFF with transition probabilities α and β [5]. The primary user's band is time slotted. $\alpha$ and $\beta$ are assumed to be the same in all bands for simplicity. The occupation probability of the band with the primary user is $P_{PB}$ that is determined by Equation (1). $P_{PI}$ can be defined by Equation (2) and represents the probability that the band not being occupied by the primary user.

$$P_{PB} = \frac{\beta}{\beta + \alpha} \quad (1)$$

$$P_{PI} = \frac{\alpha}{\alpha + \beta} \quad (2)$$

Meanwhile, we suppose there are $m$ malicious users that want to disrupt the performance of transport layer for that secondary user. The secondary user must sense the band at the beginning of the time slot to be sure that the primary user is absent. The primary users can send the information in any time slot, they want. As stated previously, the malicious users want to disturb communication of the secondary user. Thus, they wait at the beginning of time slot and sense the band to access it. If the secondary user was using the band, the malicious user would send destructive signals to destroy ACK packets.

The malicious users cooperate with each other to increase their damage on the network. It is supposed that malicious users are managed so as preventing the entrance to same band by two malicious at the same time slot. In this situation, the malicious users can occupy $m$ bands in each time slot. The strategy of the malicious users is considered simple that is randomly switching the other bands to find the secondary user.

We propose a learning method to find the best action for secondary users. Due to need for the information of other layers, we use cross layer design.

Regarding the cross layer design, we assumed that the transport layer is aware of the physical layer and act based on status of the physical layer. Figure (2) depicts the general architecture used in the proposed model.

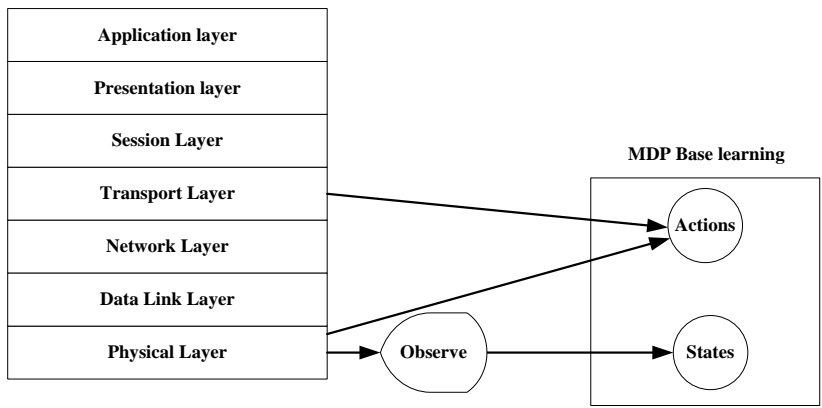

**Figure (2). The architecture of proposed model**

## IV. MDP BASED STRATEGY FOR PROPOSED MODEL

MDP is a stochastic process that models dynamics of the environment under different actions. It is a framework for representing complex multi-stage decision problems. MDP is a five-tuple $\langle S, A, P, R, \gamma \rangle$, where $S$ is the set of states, $A$ is the set of actions, $P(s'|s,a)$ is the state transition probability, $R(s', a, s)$ denotes the immediate reward values and $\gamma$ represents the discount factor which satisfies $0 \leq \gamma < 1$. In this section, the MDP model for Lion attack is explained.

### A. States

We suppose four states to model lion attack, namely: $P$, $L$, $H$ and $k$. The state of the band denoted $P$, when the primary network is present in the band. In state $P$, the secondary user can't send information and should wait until end of the time slot to find an unused band. State $L$ denotes the situation in which there is a malicious and a secondary user in the same band in absence of any primary user, results in establishing the Lion attack by malicious user. State $H$ denotes the state in which the secondary user can't receive the ACK package due to the high traffic of receiver. State $k$ is considered for the secondary user sends information successfully in absence of both primary and malicious users. $k = 1,2,3,...$ is a consecutive slot with successful transmission.

## B. Actions

The set of actions are $A = \{ST, HT, SF, HF\}$. The secondary user uses four different bisection strategies to defend against the Lion attack. The first part of strategies is related to physical layer and the second part is regard to transport layer. In each time slot, the secondary user decide whether to stay in band ($S$) or hop to another band ($H$).

Moreover, for increasing performance of the transport layer, the window size after each successful transmission doubles (maximum $K$ times); and according to TCP protocol the window size will be minimized when a heavy traffic occurs in the receiver, abbreviated by $T$. TCP protocol can use freeze algorithm to enhance throughput of the TCP protocol by fixing the window size during the congestion. The freeze algorithm is abbreviated by $F$.

In Figure (3), two actions $ST$ and $HT$ are defined for states $k$ and $H$. The first part of actions is to stay in its band or hop to other band. The second part of action ($T$) doubles the windows size for state $k$ and minimizes the windows size for state $H$. Two actions $SF$ and $HF$ are shown in Figure (4). Similarly, the first part of actions here is to stay in its band or hop from its band. The second part of actions is $F$ which freezes the widow size for both states $H$ and $L$. Figures (3) and (4) illustrate the effect of these actions on transition of states.

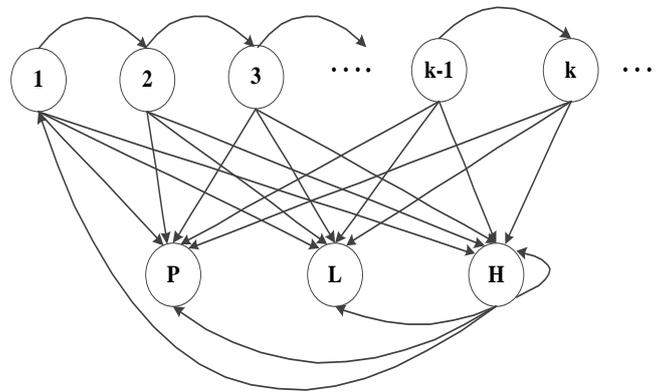

**Figure (3).the effect of two actions $ST$ and $HT$ on the transition of states**

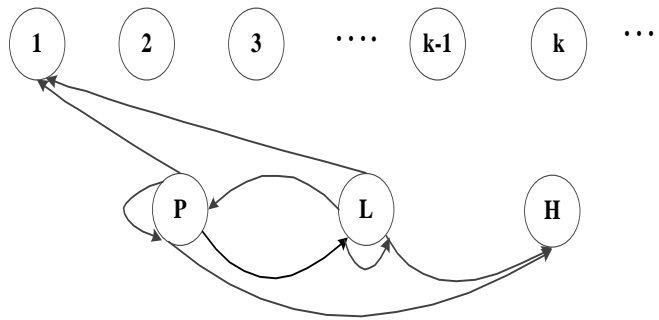

**Figure (4).the effect of two actions $SF$ and $HF$ on the transition of states**

## C. Rewards

The secondary user receives reward and penalty for each transition. $C_s$ is defined as the cost of sensing the primary users at the beginning of each time slot for the secondary user. If the secondary user wants to hop to another band, it has to change its antenna parameters and spend some energy which would incur additional costs for hopping ($C_h$). If the secondary user sends its information and the malicious users disrupt ACK package, the secondary user would lost $C_L$. When traffic of the network is heavy, the secondary users have $C_H$ penalty for unsuccessful transmission. At the end, it will get the gain $G$ when the secondary user sends its information successfully.

Taking into account the above mentioned costs and gain, rewards of the proposed model have been we calculated. When state of the band changes from $s$ to $s'$ by performing action $a$, the reward is $R(s, s', a)$. The rewards to go to state $k$ from different states by performing action $ST$ are listed in Equation (3).

$$\begin{aligned}
R(k, k+1, ST) &= G - C_s \quad k = 1, ..., \infty \\
R(k, P, ST) &= -C_s \quad k = 1, ..., \infty \\
R(k, L, ST) &= -L_L - C_s \quad k = 1, ..., \infty \\
R(k, H, ST) &= -L_H - C_s \quad k = 1, ..., \infty
\end{aligned} \quad (3)$$

Equation (4) shows the reward to go to state $H$ from different states to state $H$ with action $ST$.

$$\begin{aligned}
R(H, H, ST) &= -C_S - L_H \\
R(H, P, ST) &= -C_s \\
R(H, L, ST) &= -C_s - L_L \\
R(H, 1, ST) &= G - C_s
\end{aligned} \quad (4)$$

It is simple to calculate the expected reward for action $HT$ according to Equation (5):

$$\begin{aligned}
R(k, k+1, HT) &= G - C_s - C_h \quad k = 1, ..., \infty \\
R(k, P, HT) &= -C_s - C_h \quad k = 1, ..., K \\
R(k, H, HT) &= -L_H - C_s - C_h \quad k = 1, ..., K \\
R(k, L, HT) &= -L_L - C_s - C_h \quad k = 1, ..., K \\
R(H, H, HT) &= -C_s - C_h - C_H \\
R(H, P, HT) &= -C_s - C_h \\
R(H, L, HT) &= -C_s - L_L - C_h \\
R(H, 1, HT) &= G - C_s - C_h
\end{aligned} \quad (5)$$

Equation (6) gives the expected reward for action $SF$.

$$R(L,1,SF) = G - C_s$$
$$R(L,L,SF) = -C_s - L_L$$
$$R(L,P,SF) = -C_s \qquad (6)$$
$$R(L,H,SF) = -L_H - C_s$$
$$R(P,P,SF) = -C_s$$
$$R(P,L,SF) = -C_s - L_L$$
$$R(P,H,SF) = -L_H - C_s$$
$$R(P,1,SF) = G - C_s$$

And finally, Equation (7) presents the reward for action $HF$.

$$R(L,1,HF) = G - C_s - C_h \quad k = 1,...,\infty$$
$$R(L,L,HF) = -C_s - C_h - L_L \quad k = 1,...,\infty$$
$$R(L,P,HF) = -C_s - C_h \quad k = 1,...,\infty$$
$$R(L,H,HF) = -L_H - C_s - C_h \quad k = 1,...,\infty \qquad (7)$$
$$R(P,P,HF) = -C_s - C_h$$
$$R(P,L,HF) = -C_s - C_h - L_L$$
$$R(P,H,HF) = -L_H - C_s - C_h$$
$$R(P,1,HF) = G - C_s - C_h$$

### D. Transition Probabilities

In this section, the transition probabilities are shown for different actions. Parameter $\lambda$ is defined as the mean of the exponential distribution for heavy traffic arrival. $f_L(k)$ is the probability of occurring Lion attack that is calculated as follows:

$$f_L(k) = \begin{cases} \dfrac{m}{M - km} & k < \dfrac{M}{m} - 1 \\ 1 & otherwise \end{cases} \qquad (8)$$

Equation (9) provides the transition probabilities for action $ST$.

$$p(H|k,ST) = (1-\beta) \times \lambda \quad k = 1,...,\infty$$
$$p(P|k,ST) = \beta \quad k = 1,...,\infty$$
$$p(L|k,ST) = (1-\beta) \times (1-\lambda) \times f_L(k) \quad k = 1,...,\infty \qquad (9)$$
$$p(k+1|k,ST) = 1 - p(P|k,ST) - p(L|k,ST) - p(H|k,ST) \quad k = 1,...,\infty$$
$$p(H|H,ST) = (1-\beta) \times \lambda$$
$$p(P|H,ST) = \beta$$
$$p(L|H,ST) = (1-\beta) \times (1-\lambda) \times (\dfrac{m}{M})$$
$$p(1|H,ST) = 1 - p(H|H,ST) - p(P|H,ST) - p(L|H,ST)$$

The transition probabilities for action *HT* are shown in Equation (10):

$$p(P|k,HT) = \frac{\beta}{\alpha+\beta} \qquad k=1,...,\infty$$

$$p(H|k,HT) = (1-\frac{\beta}{\alpha+\beta}) \times \lambda \qquad k=1,...,\infty.$$

$$p(L|k,HT) = (1-\frac{\beta}{\alpha+\beta}) \times (\frac{m}{M}) \times (1-\lambda) \qquad k=1,...,\infty$$

$$p(k+1|k,HT) = 1 - p(P|k,HT) - p(H|k,HT) - p(L|k,HT) \quad k=1,...,\infty \qquad (10)$$

$$p(H|H,HT) = (1-\frac{\beta}{\alpha+\beta}) \times \lambda$$

$$p(L|H,HT) = (1-\frac{\beta}{\alpha+\beta}) \times (1-\lambda) \times (\frac{m}{M})$$

$$p(P|H,HT) = \frac{\beta}{\alpha+\beta}$$

$$p(1|H,HT) = 1 - p(P|H,HT) - p(H|H,HT) - p(L|H,HT)$$

Action *SF* has the following transition probabilities:

$$p(P|P,SF) = 1-\alpha$$

$$p(H|P,SF) = \alpha \times \lambda$$

$$p(L|P,SF) = \alpha \times (1-\lambda) \times (\frac{m}{M}) \qquad (11)$$

$$p(1|P,SF) = 1 - p(P|P,SF) - p(H|P,SF) - p(L|P,SF)$$

$$p(P|L,SF) = \beta$$

$$p(H|L,SF) = (1-\beta) \times \lambda$$

$$p(L|L,SF) = (1-\beta) \times (1-\lambda) \times (\frac{m}{M})$$

$$p(1|L,SF) = 1 - p(P|L,SF) - p(L|L,SF) - p(H|L,SF)$$

Finally, the transition probabilities for action *HF* can be obtained from Equation (12):

$$p(P|P,HF) = \frac{\beta}{\alpha+\beta}$$

$$p(H|P,HF) = \lambda \times (1-\frac{\beta}{\alpha+\beta})$$

$$p(L|P,HF) = (1-\frac{\beta}{\beta+\alpha}) \times (\frac{m}{M}) \times (1-\lambda)$$

$$p(1|P,HF) = 1 - p(L|P,HF) - p(H|P,HF) - p(P|P,HF) \qquad (12)$$

$$p(P|L,HF) = \frac{\beta}{\beta+\alpha}$$

$$p(H|L,HF) = \lambda \times (1-\frac{\beta}{\beta+\alpha})$$

$$p(L|L,HF) = (1-\frac{\beta}{\beta+\alpha}) \times (\frac{m}{M}) \times (1-\lambda)$$

$$p(1|L,HF) = 1 - p(P|L,HF) - p(H|L,HF) - p(L|L,HF)$$

### E. Solving Markov decision process

The proposed model helps the secondary user to choose the best action in each state. MDP doesn't pay attention to the expected reward but increases the long time reward for each state. The best action for each state can be calculated from Bellman Equations.

$$\pi(s) := \arg \max_{a \in ST, HT, SF, HF} \left\{ \sum_{s'} p(s', a, s) \times R(s', a, s) + \gamma \times V(s') \right\} \quad (13)$$

There are several learning methods for MDP like value iteration algorithm. Output of the value iteration algorithm is an appropriate action in each state. Figure (5) depicts learning method for the suggested MDP.

input    ST    The transtition probabilities for action Stay – TCP
           HT    The transtition probabilities for action Hop – TCP
           SF    The transtition probabilities for action Stay – Freeze
           HF    The transtition probabilities for action Hop – Freeze

           R – ST    The pay off function for action Stay – TCP
           R – HT    The pay off function for action Hop – TCP
           R – SF    The pay off function for action Stay – freeze
           R – HF    The pay off function for action Hop – freeze

Local Variable U, The payoff matrix
                U', The payoff matrix

repeat

$U \leftarrow U'$

for each state i do

$$U'(s_i) \max_{a \in ST, HT, SF, HF} \left\{ \sum_{s'} p(s', a, s) \times R(s', a, s) + \gamma \times V(s') \right\}$$

end

until $\max_{s_i} \| U(s_i) - U'(s_i) \| < \varepsilon$

return U

**Figure (5). The algorithm of proposed model**

## V. ANALYTICAL RESULTS

In this section, the analytical results derived from the proposed model are presented. We will discuss the effects of changing the amounts of m, $\beta$ and α, on the optimal strategies. When the secondary users has a constant traffic (like TV channels), change in the amounts of $\beta$ and α does not mean. But there are some networks such as WiMAX in which many kinds of traffics such as http, voice, and movie are existed. Different kinds of traffics in a network lead to the different amounts of $\beta$ and α. For example, when a primary user sends a real-time video by a WiMAX network, it uses the network continuously. But when he changes its traffic and sends a voice data, he/she probably finishes his/her communication sooner. So, if a user changes his traffic of real-time video into voice data, the amount of α will increase. We can exemplify changing the amount of $\beta$ with similar examples. The change in the amounts of $\beta$ and α result in different optimal strategies for the secondary user. So, we have discussed effects of increasing and decreasing in the amounts of $\beta$ and α on the optimal strategies that is useful for the changing-traffic networks. The initial values for parameters are assumed as follows: $m = 20$, $M = 100$, $C_s = 2$, $C_L = 10$, $C_H = 20$, $C_h = 2$, $G = 50$.

Table (1) and Figure (6) show the optimal strategies for the states H, P, L and k versus α while $\beta$ and $\lambda$ are fixed and equal to 0.5 and 0.1 respectively. The results show higher $\alpha$ cause the secondary user in the state $P$ will prefer to stay in its band, while in the states H, L and k, the secondary user prefers to hop.

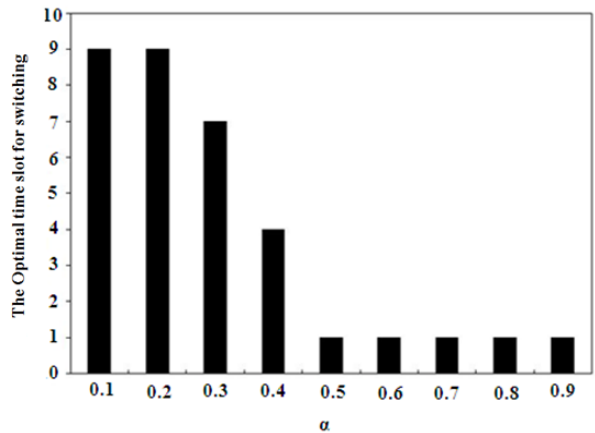

**Figure (6). The optimal time slot for switching versus α**

It is clear that only for the state $P$, both secondary and primary users are in a same band. However, in the other states namely $L$, $H$ and $k$, there is no primary users are absent in the bands occupied by the secondary users' band. According to the figure (1), in states $k$, $H$ and $L$, the changes of $\alpha$ do not have any effect on the probability of occupying the band with a primary user ($P_{PB}$) because $\beta$ is constant. Although for other bands, increasing $\alpha$ will decrease $P_{PB}$ ($P_{PB} = \frac{\beta}{\beta + \alpha}$). Thus in states $K$, $H$ and $P$, decreasing $P_{PB}$ for other bands increases the secondary user's incentive for switching.

**TABLE (1). The optimal action for secondary user versus α**

| Transition probability | The optimal strategy in physical layer | | | The optimal strategy in transport layer | | |
|---|---|---|---|---|---|---|
| α | Primary user is present | Lion attack | High Traffic | Primary user is present | Lion attack | High Traffic |
| 0.1 | Hopping | Staying | Staying | Freezing | Freezing | TCP |
| 0.2 | Hopping | Staying | Staying | Freezing | Freezing | TCP |
| 0.3 | Hopping | Staying | Staying | Freezing | Freezing | TCP |
| 0.4 | Hopping | Staying | Staying | Freezing | Freezing | TCP |
| 0.5 | Hopping | Staying | Staying | Freezing | Freezing | TCP |
| 0.6 | Staying | Staying | Staying | Freezing | Freezing | TCP |
| 0.7 | Staying | Hopping | Hopping | Freezing | Freezing | TCP |
| 0.8 | Staying | Hopping | Hopping | Freezing | Freezing | TCP |
| 0.9 | Staying | Hopping | Hopping | Freezing | Freezing | TCP |

In state $P$, increasing $\alpha$ causes the primary users leave the band with high probability as illustrated by Figure (1). Thus for the state $P$, increasing $\alpha$ leads to decrease $P_{PB}$ for the secondary user's band and consequently, the secondary user will prefer to stay in its band.

Both Table (2) and Figure (7) show the optimal strategies for the states $k$, $L$, $H$ and $P$ versus $\beta$ while $\alpha$ and $\lambda$ are fixed and equal 0.5 and 0.1 respectively. The effects of changing $\beta$ on the optimal strategies are exactly same as the effects of changing $\alpha$. The results show that by increasing α, the secondary user in state $P$ would prefer to remain in its band; and in the other state, the secondary user would prefer to hop. In state $P$, increasing $\beta$, doesn't have any effect on occupation probability of the secondary users' band with the primary users ($P_{PB}$) but the occupation probabilities of other bands will be increased ($P_{PI} = \frac{\alpha}{\alpha + \beta}$). As a result, by increasing $\beta$, the secondary user prefers to stay in its band.

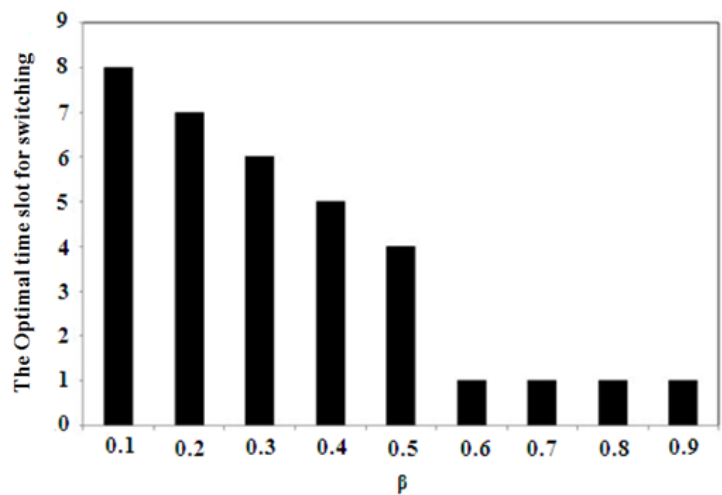

**Figure (7). The optimal time slot for switching versus β**

**TABLE (2). The optimal action for secondary user versus β**

| Transition probability | The optimal strategy in physical layer | | | The optimal strategy in transport layer | | |
|---|---|---|---|---|---|---|
| β | Primary user is present | Lion attack | High Traffic | Primary user is present | Lion attack | High Traffic |
| 0.1 | Hopping | Staying | Staying | Freezing | Freezing | TCP |
| 0.2 | Hopping | Staying | Staying | Freezing | Freezing | TCP |
| 0.3 | Hopping | Staying | Staying | Freezing | Freezing | TCP |
| 0.4 | Hopping | Staying | Staying | Freezing | Freezing | TCP |
| 0.5 | Hopping | Staying | Staying | Freezing | Freezing | TCP |
| 0.6 | Staying | Staying | Staying | Freezing | Freezing | TCP |
| 0.7 | Staying | Staying | Staying | Freezing | Freezing | TCP |
| 0.8 | Staying | Hopping | Hopping | Freezing | Freezing | TCP |
| 0.9 | Staying | Hopping | Hopping | Freezing | Freezing | TCP |

In states $L$, $H$ and $k$, the primary user doesn't exist in the secondary user's band. Figure (1) shows that in states $L$, $H$ and $k$, increasing $\beta$ results in increasing the occupation probability of the secondary user's band in the next time slot, thus the secondary user prefers to hop from its band.

In state P, increasing $\beta$ does not change the amount of $P_{PB}$, but increasing $\alpha$ result in decreasing $P_{PB}$ for the secondary's band. So as table (3) shows the secondary user prefers to stay in its band. Conversely, simultaneous increase for both $\alpha$ and $\beta$ increases $P_{PB}$ for the band in the states k, H and L. So it is rational for the secondary user to switch to other bands in those states. Table (3) and figure (8) show this fact clearly.

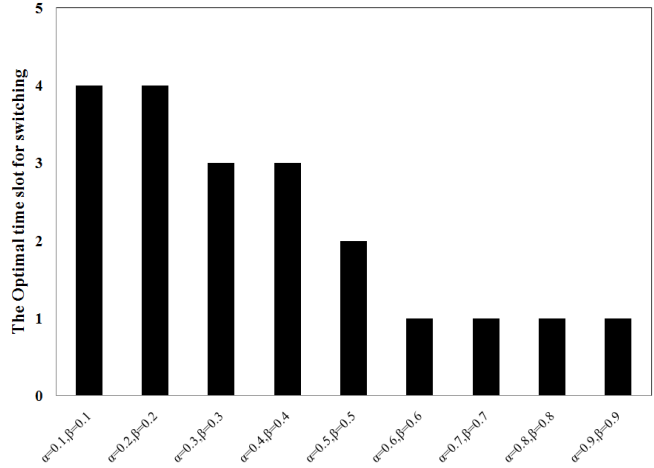

**Figure (8). The optimal time slot for switching versus α and β**

**TABLE (3). The optimal action for secondary user versus β and α**

| Transition probability | | The optimal strategy in physical layer | | | The optimal strategy in transport layer | | |
|---|---|---|---|---|---|---|---|
| α | β | Primary user is present | Lion attack | High Traffic | Primary user is present | Lion attack | High Traffic |
| 0.1 | 0.1 | Hopping | Staying | Staying | Freezing | Freezing | TCP |
| 0.2 | 0.2 | Hopping | Staying | Staying | Freezing | Freezing | TCP |
| 0.3 | 0.3 | Hopping | Staying | Staying | Freezing | Freezing | TCP |
| 0.4 | 0.4 | Hopping | Staying | Staying | Freezing | Freezing | TCP |
| 0.5 | 0.5 | Hopping | Staying | Staying | Freezing | Freezing | TCP |
| 0.6 | 0.6 | Staying | Hopping | Hopping | Freezing | Freezing | TCP |
| 0.7 | 0.7 | Staying | Hopping | Hopping | Freezing | Freezing | TCP |
| 0.8 | 0.8 | Staying | Hopping | Hopping | Freezing | Freezing | TCP |
| 0.9 | 0.9 | Staying | Hopping | Hopping | Freezing | Freezing | TCP |

The effect of increasing $\alpha$ and decreasing $\beta$ on the optimal strategies is studied in the rest of this section. Increasing $\alpha$ decreases $P_{PB}$ but changes in $\beta$ does not affect it. Also increase in $\alpha$ and decrease in $\beta$ result in decreasing $P_{PB}$ ($P_{PB} = \frac{\beta}{\beta + \alpha}$) for other bands. Although $P_{PB}$ decreases for both secondary's band and other bands but the amount of change in $P_{PB}$ for the secondary is lower than the other bands. This matter causes that the secondary prefers to switch to other bands and it is shown in table (4).

Similarly, for the states k, H and L, increase in $\alpha$ and decrease in $\beta$ result in decreasing $P_{PB}$ for both the secondary's band and other bands. Table (4) and figure (9) show that the secondary prefer to switch to other bands because of more decrease in $P_{PB}$ of other bands.

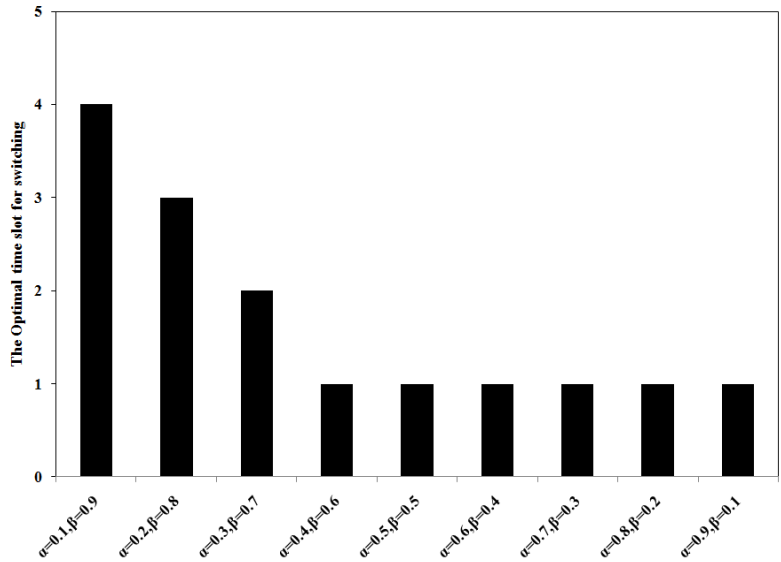

**Figure (9). The optimal time slot for switching for different values of α and β**

**TABLE (4). The optimal action for secondary user for different value of $\alpha$ and $\beta$**

| Transition probability | | The optimal strategy in physical layer | | | The optimal strategy in transport layer | | |
|---|---|---|---|---|---|---|---|
| α | β | Primary user is present | Lion attack | High Traffic | Primary user is present | Lion attack | High Traffic |
| 0.1 | 0.9 | Hopping | Hopping | Staying | Freezing | Freezing | TCP |
| 0.2 | 0.8 | Hopping | Hopping | Staying | Freezing | Freezing | TCP |
| 0.3 | 0.7 | Hopping | Hopping | Staying | Freezing | Freezing | TCP |
| 0.4 | 0.6 | Hopping | Hopping | Hopping | Freezing | Freezing | TCP |
| 0.5 | 0.5 | Hopping | Hopping | Hopping | Freezing | Freezing | TCP |
| 0.6 | 0.4 | Staying | Staying | Hopping | Freezing | Freezing | TCP |
| 0.7 | 0.3 | Staying | Staying | Hopping | Freezing | Freezing | TCP |
| 0.8 | 0.2 | Staying | Staying | Hopping | Freezing | Freezing | TCP |
| 0.9 | 0.1 | Staying | Staying | Hopping | Freezing | Freezing | TCP |

Figure (10) demonstrates that increasing the number of the malicious users, which results in more quickly switch to the secondary user to the state $k$. In states $k$, if the secondary user stays in its bands, the risk of attack is raised by increasing $m$. Thus the secondary user is better to switch as soon as possible.

Table (5) shows that the number of malicious users doesn't have any effects on the optimal strategy of the secondary user's in states $L$, $H$ and $P$. On the other hand, in state $H$, it can be inferred from Equations (10) and (11) that the probability of being attacked in the next slot by opting action HT ($p(L|H,h-tcp)$) or by selecting action ST ($p(L|H,s-tcp)$) has direct relationship with $m$ for the secondary user. Thus, increasing $m$ will equally increase the attack probability for two strategies of staying and hopping that lead to the strategies of the secondary users don't change. A similar reasoning is true for states $L$ and $P$.

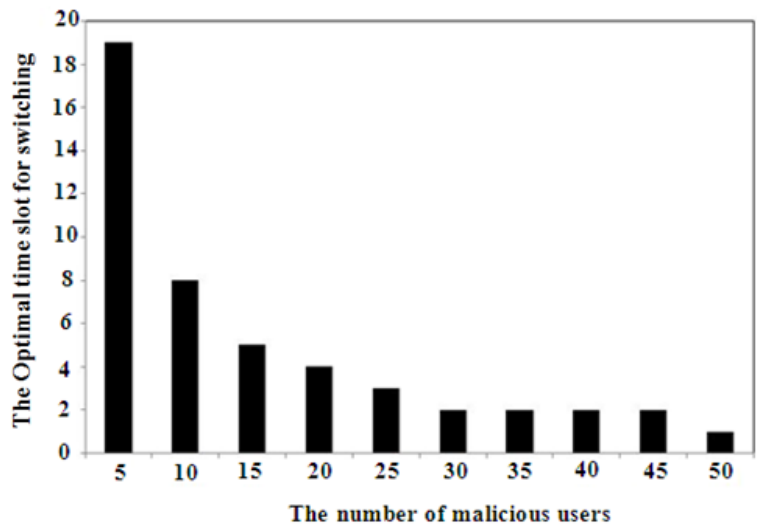

Figure (10). The optimal time slot for switching versus the number of malicious users

Table (5). The optimal action for switching versus the number of malicious users

| Number of malicious users | The optimal strategy in physical layer | | | The optimal strategy in transport layer | | |
|---|---|---|---|---|---|---|
| M | Primary user is present | Lion attack | High Traffic | Primary user is present | Lion attack | High Traffic |
| 5 | Hoping | Staying | Staying | Freezing | TCP | TCP |
| 10 | Hoping | Staying | Staying | Freezing | TCP | TCP |
| 15 | Hoping | Staying | Staying | Freezing | TCP | TCP |
| 20 | Hoping | Staying | Staying | Freezing | TCP | TCP |
| 25 | Hoping | Staying | Staying | Freezing | TCP | TCP |
| 30 | Hoping | Staying | Staying | Freezing | TCP | TCP |
| 35 | Hoping | Staying | Staying | Freezing | TCP | TCP |
| 40 | Hoping | Staying | Staying | Freezing | TCP | TCP |
| 45 | Hoping | Staying | Staying | Freezing | TCP | TCP |
| 50 | Hoping | Staying | Staying | Freezing | TCP | TCP |

# I. CONCLUSIONS

Lion attack is a multi layer attack that targets the performance of CRN try to decrease TCP connections with an opportunistic Jamming and PUE attacks.

This paper has proposed an MDP based learning algorithm for finding the best strategies in defends against Lion attack. We assume that the transport layer is aware of the physical layer's information. In such situation the main issue is selection of the appropriate strategy by both transport and physical layers. In the proposed method, the physical layer performs handoff frequency and selects either to switch or stay, while the transport layer can choose to freeze or decrease or double the length of windows size. Analytical results are discussed to find the effects of primary user behavior and the number of the malicious users, on the optimal action for the secondary user. The results show that, to increase its payoff, by increasing transition probabilities α and β, the secondary user in state $P$, should stay in its band; and in the other state, the secondary user should hop. By increasing the number of malicious users, in states $k$, the secondary user should switch as soon as possible to increase its payoff, while in states $L$, $H$ and $P$ it doesn't affect on the secondary user's optimal strategy. Future works can concentrate on implementation of the proposed model in OSI with more details and the information of data link layer can be used in cross layer design to have optimal strategy in upper layers. Also, POMDP (Partially Observable Markov Decision Process) can be used when the information of lower layer like physical layer isn't complete